\documentclass[aps,prd,reprint,twocolumn,groupedaddress,superscriptaddress,nofootinbib,longbibliography,showpacs,showkeys]{revtex4-1}
\usepackage{graphics}%
\usepackage[mathcal]{euscript}
\usepackage[latin1]{inputenc}
\usepackage{latexsym}
\usepackage{hyperref}
\usepackage{amsmath}    
\usepackage{graphicx}   
\usepackage{verbatim}  
\usepackage{color}      
\usepackage{subfigure}  
\usepackage{hyperref}   
\usepackage{bm}
\usepackage{graphicx}
\usepackage{amssymb}
\usepackage{bbm}
\usepackage{float}
\usepackage{slashed}
\usepackage{amsfonts}
\usepackage{euscript}
\usepackage{amsmath}    
\usepackage{mathrsfs} 
\usepackage{bbding}
\usepackage{pifont}
\usepackage{cancel}
\usepackage{multirow}
\usepackage{soul}

\usepackage[usenames,dvipsnames]{xcolor}
\hypersetup{colorlinks=true, citecolor=Emerald, linkcolor=Emerald,urlcolor=Emerald}

\begin{document}

\title{Topological Features of the Quantum Vacuum}

\author{Stephon Alexander}
\email{stephon\_alexander@brown.edu}
\affiliation{Brown University, Department of Physics, Providence, RI, 02912, USA}
\author{Ra\'ul Carballo-Rubio}
\email{raul.carballorubio@sissa.it}
\affiliation{SISSA, International School for Advanced Studies, Via Bonomea 265, 34136 Trieste, Italy}
\affiliation{INFN Sezione di Trieste, Via Valerio 2, 34127 Trieste, Italy} 
\begin{abstract}
A central aspect of the cosmological constant problem is to understand why vacuum energy does not gravitate. In order to account for this observation, while allowing for nontrivial dynamics of the quantum vacuum, we motivate a novel background independent theory of gravity. The theory is an extension of unimodular gravity that is described in geometric terms by means of a conformal (light-cone) structure and differential forms of degree one and two. We show that the subset of the classical field equations describing the dynamics of matter degrees of freedom and the conformal structure of spacetime are equivalent to that of unimodular gravity. The sector with vanishing matter fields and flat conformal structure is governed by the field equations of BF theory and contains topological invariants that are influenced by quantum vacuum fluctuations. Perturbative deviations from this sector lead to classical solutions that necessarily display relatively small values of the cosmological constant with respect to the would-be contribution of quantum vacuum fluctuations. This feature that goes beyond general relativity (and unimodular gravity) offers an interpretation of the smallness of the currently observed cosmological constant.

\end{abstract}

\maketitle

\def\HRULE{{\bigskip\hrule\bigskip}}


\section{Introduction}

The conformal structure of spacetime plays an essential role in the theory of general relativity. Light cones determine the causality of propagating particles and fields in any local region of spacetime. This notion of causality is also dynamical, being itself affected by matter fields. In general relativity, this geometric causal structure is embedded into the notion of a \mbox{(pseudo-)Riemannian} manifold by means of the introduction of a nowhere-vanishing differential $4$-form (the Riemannian volume form) which, together with the causal structure, determines the spacetime metric \cite{Ehlers2012}, whose dynamics is dictated by the Einstein-Hilbert action.

Not only is this conformal structure arguably more fundamental, but it has been pointed out that it is this embedding in a pseudo-Riemannian formalism the one behind the renormalization of the value of the cosmological constant in general relativity (and therefore behind one of the aspects of the cosmological constant problem). In this context, in unimodular gravity (e.g., \cite{Anderson1971,vanderBij1981,Unruh1988,Henneaux1989,Izawa1994,Alvarez2006,Smolin2010,Barcelo2014,Herrero-Valea2018}) the gravitational degrees of freedom are described by the conformal structure only, with an auxiliary nondynamical volume form. The field equations of unimodular gravity are invariant under constant shifts of the matter Lagrangian \cite{Smolin2009,Ellis2011,Ellis2013,BCRG2014}, and the effective cosmological constant is stable under radiative corrections \cite{Carballo-Rubio2015,Barcelo2015,Ardon2017}. The mechanism that guarantees the radiative stability of the cosmological constant can be traced back to the nondynamical character of the volume form. Hence, one may think that any attempt of going beyond unimodular gravity to constructing a background independent theory would not preserve the radiative stability of the cosmological constant. This paper is devoted to the analysis of this issue (see also the related works \cite{Nojiri2016,Mori2017,Nojiri2018,Mori2018}). 

Our aim here is to extend unimodular gravity to a background independent theory that still enjoys its main properties. The first question to be considered is whether it is possible to give dynamics to the nondynamical differential form in unimodular gravity so that the resulting dynamics is not equivalent to the one dictated by the Einstein-Hilbert action. To follow this path one would need a dynamical theory of differential forms that does not require a metric structure for its formulation, namely a topological field theory. One of the best known theories in this category is the so-called BF theory, described below. In this paper, we combine unimodular gravity and BF theory in order to construct a background independent theory and analyze its main properties, paying particular attention to the cosmological constant problem.

\section{The spacetime action}

\subsection{Constructing the action}

Let us start with a brief review of the relevant properties of BF theory \cite{Horowitz1989,Baez1995,Cattaneo1995,Baez1999,Gielen2010,Freidel2012,Celada2016,deGracia2017}, a topological field theory formulated on the principal bundle of a group $G$ over the spacetime manifold $\mathscr{M}$. In this principal bundle we can define a connection $\bm{A}$ and the corresponding curvature, a 2-form $\bm{F}$. To construct the action let us define a 2-form $\bm{B}$ taking values in the adjoint representation of $G$, hence displaying both spacetime and group indices. If the Lie algebra of the group $G$ is semisimple, the Killing form is nondegenerate and it can be used as an internal metric tensor in order to raise and lower group indices and to construct invariants with respect to the action of the group. In the following, $\mbox{tr}(\circ)$ will denote the trace operation on these internal indices.

With these differential forms we can construct two spacetime $4$-forms (we will be always working in $D=4$ dimensions), namely $\mbox{tr}(\bm{B}\wedge\bm{B})$ and $\mbox{tr}(\bm{B}\wedge\bm{F})$, and the action of BF theory:
\begin{equation}
\mathscr{S}_{\rm BF}=\int_{\mathscr{M}}\mbox{tr}(\bm{B}\wedge\bm{F})+\mu\,\mbox{tr}(\bm{B}\wedge\bm{B}).\label{eq:bfact}
\end{equation}
The first term plays the role of a kinetic term, while the second one is a potential term.

The idea we pursue here is to couple the differential forms in BF theory to the conformal structure of spacetime and matter fields in order to construct a background independent action. In order to to so, let us notice that the space of conformal structures is isomorphic to the space of tensor densities $|g|^{-1/4}g_{ab}$, with $g_{ab}$ a nondegenerate metric field and $g$ its determinant \cite{Ehlers2012,Bradonjic2011,Bradonjic2014}. The differential form $\mbox{tr}(\bm{B}\wedge\bm{B})$, which in coordinates takes the form
\begin{equation}
\mbox{tr}(\bm{B}\wedge\bm{B})=\sqrt{|\omega|}\,\text{d}x^0\wedge\text{d}x^1\wedge\text{d}x^2\wedge\text{d}x^3,
\end{equation}
permits to uniquely define a pseudo-Riemannian metric
\begin{equation}
\hat{g}_{ab}=\left(\frac{\omega}{g}\right)^{1/4}g_{ab}.\label{eq:metric}
\end{equation}
Let us note that the map between $\hat{g}_{ab}$ and $g_{ab}$ is not invertible, as the determinant $g$ cannot be expressed as a function of $\hat{g}_{ab}$. Hence, we are just using $\hat{g}_{ab}$ as a notational device, but we must keep in mind that it is a composite object of the gravitational field $g_{ab}$ and $\bm{B}$.

From the transformation properties of the Levi-Civita tensor one can check that
\begin{equation}
\sqrt{|\omega|}=\frac{1}{4}\epsilon^{abcd}\,\mbox{tr}(B_{ab}B_{cd})
\end{equation}
indeed transforms as expected, so that the quotient $\omega/g$ is a true scalar and $\hat{g}_{ab}$ defined in Eq. \eqref{eq:metric} a tensor field.

The Riemann curvature tensor of $\hat{g}_{ab}$ is defined as usual.  This motivates us to investigate the following background independent action:
\begin{align}
\mathscr{S}&=\int_{\mathscr{M}}\mbox{tr}(\bm{B}\wedge \bm{B})\left[\frac{1}{2\kappa}R(\hat{g})+\frac{\bar{\Lambda}}{\kappa}\right]+\int_{\mathscr{M}}\mbox{tr}(\bm{B}\wedge \bm{F})\nonumber\\
&+\mathscr{S}_{\rm M}(\hat{g},\bm{\Phi}).\label{eq:act}
\end{align}
The term, $\mathscr{S}_{\rm M}(\hat{g},\bm{\Phi})=\int_{\mathscr{M}}\mbox{tr}(\bm{B}\wedge\bm{B})\mathscr{L}_{\rm M}$ is the matter action (with the matter fields collectively denoted by $\bm{\Phi}$), minimally coupled to the metric \eqref{eq:metric}, and we have written $\mu=\bar{\Lambda}/\kappa$ for later convenience. By construction, all the terms in the action \eqref{eq:act} can be put in correspondence with a theory of a pseudo-Riemannian metric $\hat{g}_{ab}$, except for the term $\mbox{tr}(\bm{B}\wedge\bm{F})$.\footnote{Hence, it is the term giving the name to BF theory the one that breaks this equivalence, and makes the theory defined by the action \eqref{eq:act} interesting (we will see this more explicitly in the study of its equations of motion).} In other words, there is no way of constructing an invertible map between the variables $(g_{ab},\bm{B},\bm{A})$ and a new set of variables in which the first line of Eq. \eqref{eq:act} takes the form of the Einstein-Hilbert action.

\subsection{Symmetries}

The symmetries of the gravitational and matter sectors of the action are given by Weyl transformations and transverse diffeomorphisms. Weyl transformations are defined as
\begin{equation}
g_{ab}\longrightarrow \Omega^2(x) g_{ab},\label{eq:wsymdef}
\end{equation}
(note that matter fields are not affected by Weyl transformations), leaving invariant $\hat{g}_{ab}$ and therefore all the terms in the action. On the other hand, transverse diffeomorphisms are defined in a coordinate-free manner exploiting the definition of the divergence of a vector field with respect to the differential form $\mbox{tr}(\bm{B}\wedge \bm{B})$. If $\mathcal{L}_\xi$ is the Lie derivative along $\xi^a$, a vector field $\xi^a$ is transverse if and only if $\mathcal{L}_\xi(\bm{B}\wedge\bm{B})=0$. Transverse diffeomorphisms are defined as those whose generators are transverse, and can be written infinitesimally as
\begin{equation}\label{eq:symp1}
\mathcal{L}_\xi \hat{g}_{ab}=\hat{\nabla}_a\xi_b+\hat{\nabla}_b\xi_a,\qquad \hat{\nabla}_a\xi^a=0,
\end{equation}
where $\hat{\nabla}_a$ is the Levi-Civita connection associated with $\hat{g}_{ab}$. The connection $\hat{\nabla}_a$ can be also understood as an integrable Weyl connection \cite{Salim1996,Romero2012,YuanHuang2013,Barcelo2017}, as the associated connection coefficients can be written in terms of that of the Levi-Civita connection associated with $g_{ab}$, $\Gamma^c_{ab}$, as
\begin{align}\label{eq:symp2}
\hat{\Gamma}^c_{ab}&=\Gamma^c_{ab}\nonumber\\
&+\frac{1}{8}\left[\delta^c_b\partial_a\ln(\omega/g)+\delta^c_a\partial_b\ln(\omega/g)-g_{ab}\partial^c\ln(\omega/g)\right].
\end{align}
The equation above implies, in particular, that
\begin{equation}\label{eq:symp3}
\hat{\nabla}_ag_{bc}=-\frac{1}{4}g_{bc}\partial_a\ln(\omega/g).
\end{equation}
This is the usual compatibility condition of an integrable Weyl connection, which illustrates that we have arrived naturally to a particular kind of Weyl geometry. That a Weyl connection appears naturally is another manifestation of the invariance under Weyl transformations \eqref{eq:wsymdef}. Using Eqs. \eqref{eq:symp1} and  \eqref{eq:symp2}, the relevant (infinitesimal) diffeomorphisms can be alternatively written as
\begin{equation}
\mathcal{L}_\xi g_{ab}=\nabla_a\xi_b+\nabla_b\xi_a,\qquad \nabla_a\xi^a=-\frac{1}{2}\xi^a\partial_a\ln(\omega/g).
\end{equation}
On the other hand, the sector of the action involving the $\bm{B}$ and $\bm{F}$ differential forms is invariant under internal gauge transformations, given $g\in G$:
\begin{equation}
\bm{A}\rightarrow g^{-1}\bm{A}g+g^{-1}\text{d}g,\qquad \bm{B}\rightarrow g^{-1}\bm{B}g.
\end{equation}
Under these transformations, $\bm{F}\rightarrow g^{-1}\bm{F} g$ so that the two forms occurring in the action are invariant:
\begin{equation}
\mbox{tr}(\bm{F}\wedge\bm{B})\rightarrow \mbox{tr}(g^{-1}\bm{F}\wedge\bm{B}g)=\mbox{tr}(\bm{F}\wedge\bm{B}).
\end{equation}
The usual BF theory with action \eqref{eq:bfact} is invariant under the following infinitesimal transformations,
\begin{equation}
\delta\bm{A}= -2\frac{\bar{\Lambda}}{\kappa} \bm{\eta},\qquad \delta\bm{B}=\text{d}_{\bm{A}}\bm{\eta},
\end{equation}
where $\text{d}_{\bm{A}}$ is the covariant derivative associated with the connection $\bm{A}$. This symmetry renders the solutions for $\bm{B}$ in the theory wit action \eqref{eq:bfact} trivial, i.e., gauge-equivalent (locally) to the identically zero solution. However, this symmetry is broken in Eq. \eqref{eq:act} due to the couplings with the causal structure of spacetime as well as matter fields.

An additional partial symmetry that will be of relevance later is the shift of the matter Lagrangian
\begin{equation}
\mathscr{L}_{\rm M}\longrightarrow  \mathscr{L}_{\rm M}+C_0,\qquad C_0\in\mathbb{R}.\label{eq:shiftsym}
\end{equation}
This transformation leaves invariant the sector composed of matter fields and the causal spacetime structure, and modifies only the potential term in the action for the differential form $\bm{B}$. The existence of this partial symmetry is intertwined with the Weyl symmetry \eqref{eq:wsymdef}: in the absence of the latter, shifting the matter Lagrangian would not be a symmetry of this sector of the action. We will discuss in detail the meaning of this shift transformation in Secs. \ref{sec:rengroup} and \ref{sec:moresym}.

\subsection{Semiclassical renormalization group \label{sec:rengroup}}

The renormalization group of the action \eqref{eq:act} when the matter fields are quantized (defining a semiclassical theory) can be straightforwardly evaluated using standard results: the combination $\hat{g}_{ab}$ defined in Eq. \eqref{eq:metric} is a pseudo-Riemannian metric, so that all the machinery of standard techniques, such as the heat-kernel expansion, can be directly imported. Let us just quote the results that can be obtained using procedures that are thoroughly described in \cite{Birrell1982,Visser2002,Vassilevich2003,Mukhanov2007}, as explained in \cite{Carballo-Rubio2015}.

The semiclassical path integral has external fields $g_{ab}$, $\bm{B}$ and $\bm{A}$ and an integration measure that contains all the matter fields $\bm{\Phi}$, and can be evaluated using the heat-kernel expansion. Specifically, radiative semiclassical corrections affect the coupling constants $\bar{\Lambda}$ and $\kappa$ in Eq. \eqref{eq:act}. The corresponding renormalization group equations are given by
\begin{align}\label{eq:radcorr}
&\bar{\Lambda}-\bar{\Lambda}_0=C_1\alpha^4+C_2\alpha^2+C_3\ln\left(\frac{\alpha^2}{C_4}\right),\nonumber\\
&\frac{1}{\kappa}-\frac{1}{\kappa_0}=C_5\alpha^2+C_6\ln\left(\frac{\alpha^2}{C_7}\right),
\end{align}
where $\{C_i\}_{i=1}^7$ are constants with the necessary physical dimensions and $\alpha$ a cutoff. The values of these constants depend on the particle content of the matter sector (see, e.g., \cite{Visser2002} for explicit expressions). On the other hand, the bare coupling constants are given by $\bar{\Lambda}_0$ and $\kappa_0$. There also appear the usual higher-order corrections \cite{Birrell1982} which are not of particular relevance for our discussion (these will just lead to higher-derivative corrections to the trace-free gravitational field equations discussed below).

The gravitational constant $\kappa$ is renormalized as in general relativity (namely, it satisfies the same renormalization group equation). On the other hand, the coupling constant $\bar{\Lambda}$ satisfies the same renormalization group equation as the cosmological constant in general relativity. In particular, the usual arguments \cite{Martin2012} stating that the typical value of the cosmological constant in general relativity is determined by the right-hand side of the renormalization group equation in Eq. \eqref{eq:radcorr} (there is a subtlety regarding the regularization scheme \cite{Martin2012} that is nevertheless not important for our purposes), apply in this scenario to the coupling constant $\bar{\Lambda}$. This will be of importance below when studying the role that $\bar{\Lambda}$ plays in the field equations; we can anticipate, however, that $\bar{\Lambda}$ does not play the role of the cosmological constant.

\section{Some properties of the field equations}

\subsection{Classical field equations}

To obtain the field equations it is convenient to write the action \eqref{eq:act} in a coordinate basis. For example, one has
\begin{align}
&\mbox{tr}(\bm{B}\wedge\bm{B})
=\sqrt{|\omega|}\,\text{d}x^0\wedge\text{d}x^1\wedge\text{d}x^2\wedge\text{d}x^3,
\end{align}
with
\begin{equation}
\sqrt{|\omega|}=\frac{1}{4}\epsilon^{abcd}\,\mbox{tr}(B_{ab}B_{cd}).\label{eq:omegaintofB}
\end{equation}
On the other hand,
\begin{equation}
\mbox{tr}(\bm{B}\wedge\bm{F})=\frac{1}{4}\epsilon^{abcd}\,\mbox{tr}(B_{ab}F_{cd})\,\text{d}x^0\wedge\text{d}x^1\wedge\text{d}x^2\wedge\text{d}x^3.
\end{equation}
The variation with respect to $B_{ab}$ leads then to
\begin{align}
&\frac{1}{4}\epsilon^{abcd}F^I_{cd}\nonumber\\
&+\frac{\delta}{\delta B^I_{ab}}\left\{\int_{\mathscr{M}}\mbox{tr}(\bm{B}\wedge \bm{B})\left[\frac{1}{2\kappa}R(\hat{g})+\frac{\bar{\Lambda}}{\kappa}\right]+\mathscr{S}_{\rm M}(\hat{g},\bm{\Phi})\right\}=0.\label{eq:varB1}
\end{align}
The metric field $\hat{g}_{ab}$ is a nontrivial function of the fields $\bm{B}$ and $g_{ab}$. The variation of the action with respect these two fields can be evaluated using the chain rule, considering first variations $\delta\hat{g}^{ab}$ and expressing these in terms of $\delta B^I_{ab}$ and $\delta g^{ab}$. Under a general variation $\delta\hat{g}^{ab}$,
\begin{align}
&\delta \int_{\mathscr{M}}\mbox{tr}(\bm{B}\wedge \bm{B})R(\hat{g})\nonumber\\
&=\int_{\mathscr{M}}\text{d}^4x\sqrt{|\omega|}\left[R_{cd}(\hat{g})-\frac{1}{2}R(\hat{g})\hat{g}_{cd}\right]\delta\hat{g}^{cd}.
\end{align}
Then, the second term in Eq. \eqref{eq:varB1} is proportional to the trace of the Einstein field equations (with cosmological constant $-\bar{\Lambda}$) evaluated on the composite metric $\hat{g}_{ab}$:
\begin{align}
&\frac{\delta}{\delta B^I_{ab}}\left\{\int_{\mathscr{M}}\mbox{tr}(\bm{B}\wedge \bm{B})\left[\frac{1}{2\kappa}R(\hat{g})+\bar{\Lambda}\right]+\mathscr{S}_{\rm M}(\hat{g},\bm{\Phi})\right\}\nonumber\\
&=\frac{\sqrt{|\omega|}}{2\kappa}\left[R_{cd}(\hat{g})-\frac{1}{2}R(\hat{g})\hat{g}_{cd}-\bar{\Lambda}\hat{g}_{cd}-\kappa T_{cd}\right]\frac{\delta\hat{g}^{cd}}{\delta B^I_{ab}}\nonumber\\
&=-\frac{\sqrt{|\omega|}}{8\kappa}\left[G_{cd}(\hat{g})-\bar{\Lambda}\hat{g}_{cd}-\kappa T_{cd}\right]\frac{|g|^{1/4}}{|\omega|^{5/4}}g^{cd}\frac{\delta|\omega|}{\delta B^I_{ab}}\nonumber\\
&=\frac{1}{8 \kappa}\left[R(\hat{g})+4\bar{\Lambda}+\kappa T\right]\epsilon^{abcd}B^I_{cd}.
\end{align}
We have used $\hat{g}^{ab}=(\omega/g)^{-1/4}g^{ab}$ and
\begin{equation}
\frac{\delta|\omega|}{\delta B^I_{ab}}=\sqrt{|\omega|}\,\epsilon^{abcd}B^I_{cd},
\end{equation}
which can be obtained from Eq. \eqref{eq:omegaintofB}. $G_{ab}$ is the usual Einstein tensor, and we follow the standard definition of the Belinfante-Rosenfeld stress-energy tensor of matter fields:
\begin{equation}
T_{ab}=-\frac{2}{\sqrt{|\omega|}}\frac{\delta\mathscr{S}_{\rm M}(\hat{g},\bm{\Phi})}{\delta\hat{g}^{ab}}=-2\frac{\partial\mathscr{L}_{\rm M}}{\partial \hat{g}^{ab}}+\mathscr{L}_{\rm M}\hat{g}_{ab}.
\end{equation}
Using the relations above, Eq. \eqref{eq:varB1} can be written as
\begin{equation}
\bm{F}+\frac{1}{2\kappa}[4\bar{\Lambda}+R(\hat{g})+\kappa T]\bm{B}=0.\label{eq:eqmot3}
\end{equation}
Variations of the action with respect to $\bm{A}$ yield simply
\begin{equation}
\text{d}_{\bm{A}}\bm{B}=0.\label{eq:eqmot1}
\end{equation}
Lastly, variations with respect to $g^{ab}$ lead to the trace-free Einstein field equations:
\begin{align}
&\frac{\delta}{\delta g^{ab}}\left\{\int_{\mathscr{M}}\mbox{tr}(\bm{B}\wedge \bm{B})\left[\frac{1}{2\kappa}R(\hat{g})+\frac{\bar{\Lambda}}{\kappa}\right]+\mathscr{S}_{\rm M}(\hat{g},\bm{\Phi})\right\}\nonumber\\
&=\frac{\sqrt{|\omega|}}{2\kappa}\left[G_{cd}(\hat{g})-\bar{\Lambda}\hat{g}_{cd}-\kappa T_{cd}\right]\frac{\delta\hat{g}^{cd}}{\delta g^{ab}}\nonumber\\
&=\frac{\sqrt{|\omega|}}{2\kappa}\left[G_{cd}(\hat{g})-\bar{\Lambda}\hat{g}_{cd}-\kappa T_{cd}\right]\frac{|g|^{1/4}}{|\omega|^{1/4}}\left(\delta^c_a\delta^d_b-\frac{1}{4}g^{cd}g_{ab}\right)\nonumber\\
&=\frac{\sqrt{|\omega|}}{2\kappa}\left(g/\omega\right)^{1/4}\nonumber\\
&\times\left[R_{ab}(\hat{g})-\frac{1}{4}R(\hat{g})\hat{g}_{ab}-\kappa\left(T_{ab}-\frac{1}{4}T\hat{g}_{ab}\right)\right].
\end{align}
The gravitational field equations are therefore given by
\begin{equation}
R_{ab}(\hat{g})-\frac{1}{4}R(\hat{g})\hat{g}_{ab}=\kappa\left(T_{ab}-\frac{1}{4}T\hat{g}_{ab}\right).\label{eq:eqmot2}
\end{equation}
Eqs. \eqref{eq:eqmot3}, \eqref{eq:eqmot1} and \eqref{eq:eqmot2} are the field equations of the theory.

\subsection{Comparison with unimodular gravity and general relativity \label{sec:moresym}}

The field equations describe a generalization of unimodular gravity, with two new equations, \eqref{eq:eqmot3} and \eqref{eq:eqmot1}. In this section we explain the implications of this extension, highlighting the similarities and differences with both unimodular gravity and general relativity.

An important property of unimodular gravity that makes it different from general relativity is the invariance of Eq. \eqref{eq:eqmot2} under the shift
\begin{equation}
T_{ab}\longrightarrow T_{ab}+C_0g_{ab},\label{eq:ttrans}
\end{equation}
which is nothing but the shift symmetry \eqref{eq:shiftsym} when expressed in terms of the stress-energy tensor of matter. An additional symmetry of unimodular gravity is
\begin{equation}
R_{ab}\longrightarrow R_{ab}+\kappa D_0g_{ab}.\label{eq:rtrans}
\end{equation}
In general relativity, these two transformations are not symmetries of the classical field equations independently, but only their combination with $D_0=-C_0$. This is another way of stating the cosmological constant problem in general relativity: quantum vacuum fluctuations leading to a nonzero $C_0$ produce spacetime curvatures that are constrained to be proportional to $D_0=-C_0$. 

However, in unimodular gravity the constants $C_0$ and $D_0$ in the symmetry transformations \eqref{eq:ttrans} and \eqref{eq:rtrans} are unrelated. These two independent symmetries are preserved in the generalization of unimodular gravity that we are considering there. The transformations that extend Eqs. \eqref{eq:ttrans} and \eqref{eq:rtrans} are given, respectively, by
\begin{equation}
T_{ab}\longrightarrow T_{ab}+C_0g_{ab},\qquad F_{ab}^I\longrightarrow F_{ab}^I-2C_0B_{ab}^I\label{eq:ttrans2},
\end{equation}
and
\begin{equation}
R_{ab}\longrightarrow R_{ab}+\kappa D_0g_{ab},\qquad F_{ab}^I\longrightarrow F_{ab}^I-2D_0B_{ab}^I.\label{eq:rtrans2}
\end{equation}
Hence, as in unimodular gravity, zero-point shifts of the spacetime curvature and of the stress-energy tensor of matter can be performed independently, so that there is no link between these physical notions.

In general relativity, the stress-energy tensor of matter is identically conserved due to diffeomorphism invariance. We can show that a similar statement is valid for the theory introduced in this paper. First of all, let us note that Eqs. \eqref{eq:eqmot3} and \eqref{eq:eqmot1} together imply the constraint
\begin{equation}
\text{d}[R(\hat{g})+\kappa T]\wedge\bm{B}=0.\label{eq:noncons1}
\end{equation}
Let us define the exact one-form $\text{d}S$ as
\begin{equation}
\frac{1}{4}\text{d}[R(\hat{g})+\kappa T]=\kappa\, \text{d}S,
\end{equation}
so that Eq. \eqref{eq:noncons1} is equivalent to
\begin{equation}
\text{d}S\wedge\bm{B}=0.
\end{equation}
The interpretation of the one-form $\text{d}S$ is the following. Taking the divergence with respect to the Levi-Civita connection of $\hat{g}_{ab}$ in Eq. \eqref{eq:eqmot2}, and using the Bianchi identities satisfied by the Ricci tensor, it follows that the divergence of the stress-energy tensor is given by
\begin{equation}
\hat{\nabla}^bT_{ab}=\frac{1}{4\kappa }\partial_a[R(\hat{g})+\kappa T]=\partial_a S,
\end{equation}
so that a nonzero $S$ would describe the violation of energy conservation \cite{Josset2016,Perez2017,Perez2018}. We have to determine whether there are classical solutions for which $\text{d}S\neq0$. However, as explained in App. \ref{sec:app}, solutions with $|\omega|\neq0$ necessarily satisfy $\text{d}S=0$. Therefore, it has to be
\begin{equation}
\hat{\nabla}^bT_{ab}=0.\label{eq:cons}
\end{equation}
It follows that the gravitational field equations \eqref{eq:eqmot2} can be written as
\begin{equation}
R_{ab}(\hat{g})-\frac{1}{2}R(\hat{g})\hat{g}_{ab}+\Lambda\hat{g}_{ab}=\kappa T_{ab},\label{eq:intgraveqs}
\end{equation}
where $\Lambda$ is an integration constant that plays the role of the cosmological constant. These equations are manifestly invariant under Weyl transformations \eqref{eq:wsymdef}. 

In general relativity, the cosmological constant is a coupling constant and therefore has a fixed value up to the renormalization group flow. However, $\Lambda$ in Eq. \eqref{eq:intgraveqs} is an integration constant that does not have a definite value until a given solution is chosen, and that is not affected by radiative corrections that modify instead $\bar{\Lambda}$, as made explicit in Eq. \eqref{eq:radcorr}. Its value is shifted by the transformations \eqref{eq:ttrans2} and \eqref{eq:rtrans2} as
\begin{equation}
\Lambda\longrightarrow \Lambda+D_0+\kappa C_0.
\end{equation}
It is also worth stressing that Eq. \eqref{eq:intgraveqs} contains no trace of the coupling constant $\bar{\Lambda}$, which in this theory receives the radiative corrections that in general relativity modify the value of the cosmological constant.

The form of the field equations can be simplified taking into account that conservation of the stress-energy tensor of matter or, equivalently, Eq. \eqref{eq:cons}, implies that Eq. \eqref{eq:eqmot3} can be written as
\begin{equation}
\bm{F}+\frac{2}{\kappa}(\bar{\Lambda}+\Lambda)\bm{B}=0.\label{eq:eqmot4}
\end{equation}
We see that the coupling constant $\bar{\Lambda}$ shows up as a constant source to the curvature $\bm{F}$ of the connection $\bm{A}$. Unless the group $G$ is abelian, Eqs. \eqref{eq:eqmot1} and \eqref{eq:eqmot4} have to be solved for $\bm{A}$ and $\bm{B}$ simultaneously, and then Eq. \eqref{eq:intgraveqs} can be used to determine the gravitational field $g_{ab}$.

Hence, we have two constants $\bar{\Lambda}$ and $\Lambda$ that act as constant sources of curvature, although for different curvatures ($F^I_{ab}$ and $R_{ab}$, respectively). Let us recall that, as explained in Sec. \ref{sec:rengroup}, the quantity that is renormalized due to the fluctuations of the quantum vacuum is $\bar{\Lambda}$. On the other hand, the effective cosmological constant $\Lambda$ is an independent quantity, the value of which depends on the local properties of matter and the gravitational field (it is an integration constant that is determined by the initial state of these fields). There is no renormalization group for $\Lambda$; in other words, it is not renormalized by quantum corrections. The splitting $\bar{\Lambda}+\Lambda$ in Eq. \eqref{eq:eqmot4} can be seen as corresponding to the quantum and classical parts, respectively, of what in general relativity is understood as the cosmological constant that acts as a source of the curvature $R_{ab}$. However, here these two parts have different dynamical effects: only $\Lambda$ acts as a true cosmological constant, sourcing the curvature $R_{ab}$, while the quantum contribution $\bar{\Lambda}$ is funneled to the BF sector of the theory, sourcing only the curvature $F^I_{ab}$. This can be alternatively seen by studying particular solutions, as we do in the following.

\subsection{Vacuum solutions with flat conformal structure \label{sec:qvdef}}

Let us first consider the solutions of the field equations for field configurations with no local metric structure. That is, we will consider test matter fields and zero Riemann tensor; a nonzero Riemann tensor can be locally detected by using trajectories of test particles over lengths of the order of the (inverse) associated curvature. One has then
\begin{equation}
R(\hat{g})=\kappa\,T=0,
\end{equation}
leaving only the field equations of BF theory \eqref{eq:bfact},
\begin{equation}
\bm{F}+2\frac{\bar{\Lambda}}{\kappa} \bm{B}=0,\qquad \text{d}_{\bm{A}}\bm{B}=0.
\end{equation}
Hence, in the absence of local physical excitations the theory still displays nontrivial dynamics that describes additional properties of the manifold that is locally Minkowskian. In semiclassical general relativity this sector of the theory is taken as the one describing the quantum vacuum, void of excitations of the matter fields. However, in general relativity the quantum vacuum energy renormalizes the value of the cosmological constant, and drives the quantum vacuum away from $\Lambda=0$.

Due to the similarity with general relativity, we consider in the following that this sector (or truncation) of the theory, the dynamics of which is described by BF theory, provides an effective description of the dynamical properties of the quantum vacuum. A first indication that this definition of the quantum vacuum is reasonable is that, as explained in Sec. \ref{sec:rengroup}, quantum vacuum effects renormalize the coupling constant $\bar{\Lambda}$, thus modifying the state of the connection $A^I_a$ associated with the quantum vacuum. Hence, all the solutions in this sector have $\Lambda=0$ but $\bar{\Lambda}\neq0$.

Let us consider a particular solution of the field equations for the abelian case $G=\mbox{U}(1)$. The metric tensor and $B_{ab}$ form are given by
\begin{equation}
\eta_{ab}=\left(\begin{array}{cccc}-1 &0&0&0\\ 0&1&0&0\\0&0&1&0\\0&0&0&1\end{array} \right),\qquad B_{ab}=\frac{1}{2}\left(\begin{array}{cccc}0 &1&0&0\\ -1&0&0&0\\0&0&0&1\\0&0&-1&0\end{array} \right),
\end{equation}
while the associated curvature form is given by
\begin{equation}
F_{ab}=\frac{\bar{\Lambda}}{\kappa}\left(\begin{array}{cccc}0 &-1&0&0\\ 1&0&0&0\\0&0&0&-1\\0&0&1&0\end{array} \right).
\end{equation}
A connection leading to this curvature is for instance
\begin{equation}
A=\frac{\bar{\Lambda}}{\kappa}\left(x^1\,\text{d}x^0+x^3\,\text{d}x^2\right).
\end{equation}
We see that the curvature $F_{ab}$ is of the order of the coupling constant $\bar{\Lambda}/\kappa$. Hence, this solution is characterized by vanishing metric curvature but nonzero curvature $F_{ab}$ proportional to $\bar{\Lambda}$ which, in turn, would be of the order of magnitude determined by quantum vacuum fluctuations (due to the renormalization group equation discussed in Sec. \ref{sec:rengroup}). We can consider for instance the topological invariant given by the integral of the Chern-Weil form,
\begin{equation}
\int_{\mathscr{M}}\mbox{tr}(\bm{F}\wedge\bm{F})=4\frac{\bar{\Lambda}^2}{\kappa^2}\mbox{Vol}(\mathscr{M}),\label{eq:qvobs1}
\end{equation}
where we have defined $\mbox{Vol}(\mathscr{M})=\int_{\mathscr{M}}\mbox{tr}(\bm{B}\wedge\bm{B})$. Equivalently, we can write
\begin{equation}
\frac{1}{\mbox{Vol}(\mathscr{M})}\int_{\mathscr{M}}\mbox{tr}(\bm{F}\wedge\bm{F})=4\frac{\bar{\Lambda}^2}{\kappa^2}.\label{eq:qvobs2}
\end{equation}
We see that the value of this invariant is determined by the fluctuations of the quantum vacuum through the first relation in Eq. \eqref{eq:radcorr}. As discussed in detail in \cite{Martin2012}, one has for the particle spectrum of the Standard Model that \mbox{$\bar{\Lambda}\sim 10^{8}\mbox{ GeV}^4$}.

\subsection{Perturbations of the quantum vacuum}

The next step we want to describe is the introduction of local metric structures that are perturbations of the sector of solutions described just above. The introduction of local metric structures induces generally a nonzero cosmological constant $\Lambda$. We can exploit the existence of invariants such as Eq. \eqref{eq:qvobs2} in order to provide a definition of what `perturbative' means when $\Lambda\neq0$. Using Eq. \eqref{eq:eqmot4} we can write in general
\begin{equation}\label{eq:vacpert}
\frac{1}{\mbox{Vol}(\mathscr{M})}\int_{\mathscr{M}}\mbox{tr}(\bm{F}\wedge\bm{F})=\frac{4}{\kappa^2}(\bar{\Lambda}+\Lambda)^2.
\end{equation}
Perturbative deviations are characterized by 
\begin{equation}
\bm{F}+\delta\bm{F},\qquad \delta \bm{F}\ll \bm{F}.\label{eq:fpert}
\end{equation}
In particular, Eq. \eqref{eq:qvobs2} must be modified only perturbatively. This implies that $\bar{\Lambda}$ plays the role of the scale to be perturbed (which is reasonable as it is the only dimensional constant that plays a role in the description of the vacuum). We can read from Eq. \eqref{eq:vacpert} that perturbative deviations from the vacuum must therefore satisfy
\begin{equation}\label{eq:ineq1}
\Lambda\ll \bar{\Lambda}.
\end{equation}
We can therefore define a special subset of classical solutions, in the following way: a given solution of the field equations with $\Lambda\neq0$ is a perturbation of the vacuum with $\Lambda=0$ if and only if the effective cosmological constant $\Lambda$ is small with respect to the scale $\bar{\Lambda}$. 

While the typical value of $\bar{\Lambda}$ is determined by the fluctuations of the quantum vacuum, in order to fix the value of the integration constant $\Lambda$ we have to resort to observations. The observational value satisfies \cite{Martin2012}
\begin{equation}\label{eq:hier}
\frac{\Lambda}{\bar{\Lambda}}\sim 10^{-55}\lll1.
\end{equation}
It follows that, in this framework, the current state of the universe would indeed be given by one of these solutions satisfying Eq. \eqref{eq:ineq1}. We can reverse this argument, thus concluding that Eq. \eqref{eq:hier} is just a manifestation of the fact that the universe is extremely close to the vacuum solution described in Sec. \ref{sec:qvdef}, representing the quantum vacuum. The mismatch between these two scales finds then this natural interpretation in the framework being described. These considerations are remarkably similar to Volovik's proposal to understand the value of the cosmological constant as arising from a small displacement with respect to (or perturbative deviation from) the equilibrium state of the universe \cite{Volovik2003,Volovik2004,Volovik2006,Klinkhamer2008,Klinkhamer2008b,Klinkhamer2009}. In fact, the present formalism could be understood as an explicit realization of these ideas.

Regarding previous work, it is worth stressing that the present formalism is different from scenarios in which the cosmological constant is a dynamical variable (e.g., \cite{Alexander2018}) and also from the sequestering mechanism introduced in \cite{Kaloper2013,Kaloper2014} (see also \cite{Tsukamoto2017}). The sequestering mechanism is more restrictive regarding the value of the cosmological constant, as it fixes the latter in terms of a cosmological average even in the classical theory (the introduction of this global constraint is at the core of this mechanism). We can discriminate between two elements (or two cosmological problems): (i) radiative stability of the cosmological constant (this is guaranteed in both our case and sequestering) and (ii) fixing a particular value of the cosmological constant. In our case, the classical theory does not fix the value of the cosmological constant, although there are hints that unimodular gravity should lead to similar cosmological averages as the sequestering mechanism when the quantization of the gravitational field is carried out \cite{Smolin2009}. Due to its relation to unimodular gravity, it would be certainly interesting to study this aspect in the theory introduced here. This is, however, out of the scope of this work.

Another difference with respect to the sequestering mechanism arises when considering the effect of phase transitions. We have stressed above that, in the framework discussed in this paper, the cosmological constant is a radiatively stable (and in fact, invariant) parameter that must be fixed by matching with observations, similarly to other fundamental constants such as the electron charge. The present-day value of the cosmological constant leads to Eq. \eqref{eq:hier} which, in our framework, is naturally interpreted as the current state of the universe representing a perturbative deviation with respect to the quantum vacuum. However, cosmological phase transitions do affect the value of the cosmological constant in this scenario, and therefore may drive the universe out of this near-equilibrium when going backwards in time. From the discussion in \cite{Martin2012} for instance, it is straightforward to check that the cosmological QCD transition preserves the near-equilibrium condition but the electroweak phase transition does not. Hence, in this theory the state of the universe before the latter phase transition cannot correspond to a perturbative deviation with respect to the vacuum. Let us stress that, as already emphasized by Weinberg \cite{Weinberg1989}, there is no contradiction at all between known astrophysical observations and a cosmological constant that grows in the past due to phase transitions.

\section{Conclusions}
 
Using the well-known BF theory, we have extended unimodular gravity to a background independent theory that displays a number of interesting properties. Our main conclusions are: (1) it is possible to construct background independent theories with dynamical volume forms that still display the main properties of the trace-free Einstein field equations, (2) this entails using and merging together tools of topological field theory and Weyl geometry, and (3) the new features beyond unimodular gravity offer a natural interpretation of the smallness of the observed cosmological constant when compared to the would-be contribution of the quantum vacuum (although not a prediction of its value). For future work it would be interesting to study the cosmological and astrophysical implications of this theory, and also whether it may be possible to naturally get a dark sector in this framework.

\acknowledgments

The authors are grateful to Evan McDonough and Lee Smolin for useful discussions. R.C-R. gratefully acknowledges the hospitality of Brown University and Perimeter Institute, where part of this work was completed. This research was supported in part by Perimeter Institute for Theoretical Physics. Research at Perimeter Institute is supported by the Government of Canada through the Department of Innovation, Science and Economic Development and by the Province of Ontario through
the Ministry of Research and Innovation.

\vspace{10pt}
\appendix
\section{Some basic algebra \label{sec:app}}

For simplicity, let us consider $G=\rm{U}(1)$ in this appendix. In $D=4$ dimensions, the determinant of an antisymmetric matrix verifies \cite{Ledermann1993}
\begin{equation}
\mbox{det}(B_{ab})=\left(\frac{1}{4}\epsilon^{abcd}B_{ab}B_{cd}\right)^2=|\omega|.
\end{equation}
Hence, it follows that a nondegenerate metric structure demands that $\mbox{det}(B_{ab})\neq0$.

Let us now consider a one-form $A$ and analyze the content of the equation $A\wedge B=0$. If at least one component of $A$ is nonzero, which we initially take to be $A_0$ without loss of generality, then only three of these equations are independent:
\begin{align}
B_{12}=\frac{A_1B_{02}-A_2B_{01}}{A_0},\nonumber\\
B_{13}=\frac{A_1B_{03}-A_3B_{01}}{A_0},\nonumber\\
B_{23}=\frac{A_2B_{03}-A_3B_{02}}{A_0}.
\end{align}
It is straightforward to show that, if these three conditions are satisfied, then $\mbox{det}(B_{ab})=0$. Hence, it must be $A_0=0$. We can then apply the same argument to the remaining components of $A$ recursively, thus concluding that $A=0$ as long as $\mbox{det}(B_{ab})\neq0$. 

\raggedright
\bibliography{refs}	

\begin{thebibliography}{55}%
\makeatletter
\providecommand \@ifxundefined [1]{%
 \@ifx{#1\undefined}
}%
\providecommand \@ifnum [1]{%
 \ifnum #1\expandafter \@firstoftwo
 \else \expandafter \@secondoftwo
 \fi
}%
\providecommand \@ifx [1]{%
 \ifx #1\expandafter \@firstoftwo
 \else \expandafter \@secondoftwo
 \fi
}%
\providecommand \natexlab [1]{#1}%
\providecommand \enquote  [1]{``#1''}%
\providecommand \bibnamefont  [1]{#1}%
\providecommand \bibfnamefont [1]{#1}%
\providecommand \citenamefont [1]{#1}%
\providecommand \href@noop [0]{\@secondoftwo}%
\providecommand \href [0]{\begingroup \@sanitize@url \@href}%
\providecommand \@href[1]{\@@startlink{#1}\@@href}%
\providecommand \@@href[1]{\endgroup#1\@@endlink}%
\providecommand \@sanitize@url [0]{\catcode `\\12\catcode `\$12\catcode
  `\&12\catcode `\#12\catcode `\^12\catcode `\_12\catcode `\%12\relax}%
\providecommand \@@startlink[1]{}%
\providecommand \@@endlink[0]{}%
\providecommand \url  [0]{\begingroup\@sanitize@url \@url }%
\providecommand \@url [1]{\endgroup\@href {#1}{\urlprefix }}%
\providecommand \urlprefix  [0]{URL }%
\providecommand \Eprint [0]{\href }%
\providecommand \doibase [0]{http://dx.doi.org/}%
\providecommand \selectlanguage [0]{\@gobble}%
\providecommand \bibinfo  [0]{\@secondoftwo}%
\providecommand \bibfield  [0]{\@secondoftwo}%
\providecommand \translation [1]{[#1]}%
\providecommand \BibitemOpen [0]{}%
\providecommand \bibitemStop [0]{}%
\providecommand \bibitemNoStop [0]{.\EOS\space}%
\providecommand \EOS [0]{\spacefactor3000\relax}%
\providecommand \BibitemShut  [1]{\csname bibitem#1\endcsname}%
\let\auto@bib@innerbib\@empty
\bibitem [{\citenamefont {Ehlers}\ \emph {et~al.}(2012)\citenamefont {Ehlers},
  \citenamefont {Pirani},\ and\ \citenamefont {Schild}}]{Ehlers2012}%
  \BibitemOpen
  \bibfield  {author} {\bibinfo {author} {\bibfnamefont {J.}~\bibnamefont
  {Ehlers}}, \bibinfo {author} {\bibfnamefont {F.~A.~E.}\ \bibnamefont
  {Pirani}}, \ and\ \bibinfo {author} {\bibfnamefont {A.}~\bibnamefont
  {Schild}},\ }\bibfield  {title} {\enquote {\bibinfo {title} {Republication
  of: The geometry of free fall and light propagation},}\ }\href {\doibase
  10.1007/s10714-012-1353-4} {\bibfield  {journal} {\bibinfo  {journal}
  {General Relativity and Gravitation}\ }\textbf {\bibinfo {volume} {44}},\
  \bibinfo {pages} {1587--1609} (\bibinfo {year} {2012})}\BibitemShut {NoStop}%
\bibitem [{\citenamefont {Anderson}\ and\ \citenamefont
  {Finkelstein}(1971)}]{Anderson1971}%
  \BibitemOpen
  \bibfield  {author} {\bibinfo {author} {\bibfnamefont {J.~L.}\ \bibnamefont
  {Anderson}}\ and\ \bibinfo {author} {\bibfnamefont {D.}~\bibnamefont
  {Finkelstein}},\ }\bibfield  {title} {\enquote {\bibinfo {title}
  {{Cosmological constant and fundamental length}},}\ }\href {\doibase
  10.1119/1.1986321} {\bibfield  {journal} {\bibinfo  {journal} {Am. J. Phys.}\
  }\textbf {\bibinfo {volume} {39}},\ \bibinfo {pages} {901--904} (\bibinfo
  {year} {1971})}\BibitemShut {NoStop}%
\bibitem [{\citenamefont {van~der Bij}\ \emph {et~al.}(1982)\citenamefont
  {van~der Bij}, \citenamefont {van Dam},\ and\ \citenamefont
  {Ng}}]{vanderBij1981}%
  \BibitemOpen
  \bibfield  {author} {\bibinfo {author} {\bibfnamefont {J.~J.}\ \bibnamefont
  {van~der Bij}}, \bibinfo {author} {\bibfnamefont {H.}~\bibnamefont {van
  Dam}}, \ and\ \bibinfo {author} {\bibfnamefont {Y.~J.}\ \bibnamefont {Ng}},\
  }\bibfield  {title} {\enquote {\bibinfo {title} {{The Exchange of Massless
  Spin Two Particles}},}\ }\href {\doibase 10.1016/0378-4371(82)90247-3}
  {\bibfield  {journal} {\bibinfo  {journal} {Physica}\ }\textbf {\bibinfo
  {volume} {116A}},\ \bibinfo {pages} {307--320} (\bibinfo {year}
  {1982})}\BibitemShut {NoStop}%
\bibitem [{\citenamefont {Unruh}(1989)}]{Unruh1988}%
  \BibitemOpen
  \bibfield  {author} {\bibinfo {author} {\bibfnamefont {W.~G.}\ \bibnamefont
  {Unruh}},\ }\bibfield  {title} {\enquote {\bibinfo {title} {{A Unimodular
  Theory of Canonical Quantum Gravity}},}\ }\href {\doibase
  10.1103/PhysRevD.40.1048} {\bibfield  {journal} {\bibinfo  {journal} {Phys.
  Rev.}\ }\textbf {\bibinfo {volume} {D40}},\ \bibinfo {pages} {1048} (\bibinfo
  {year} {1989})}\BibitemShut {NoStop}%
\bibitem [{\citenamefont {Henneaux}\ and\ \citenamefont
  {Teitelboim}(1989)}]{Henneaux1989}%
  \BibitemOpen
  \bibfield  {author} {\bibinfo {author} {\bibfnamefont {M.}~\bibnamefont
  {Henneaux}}\ and\ \bibinfo {author} {\bibfnamefont {C.}~\bibnamefont
  {Teitelboim}},\ }\bibfield  {title} {\enquote {\bibinfo {title} {{The
  Cosmological Constant and General Covariance}},}\ }\href {\doibase
  10.1016/0370-2693(89)91251-3} {\bibfield  {journal} {\bibinfo  {journal}
  {Phys. Lett.}\ }\textbf {\bibinfo {volume} {B222}},\ \bibinfo {pages}
  {195--199} (\bibinfo {year} {1989})}\BibitemShut {NoStop}%
\bibitem [{\citenamefont {Izawa}(1995)}]{Izawa1994}%
  \BibitemOpen
  \bibfield  {author} {\bibinfo {author} {\bibfnamefont {K.~I.}\ \bibnamefont
  {Izawa}},\ }\bibfield  {title} {\enquote {\bibinfo {title} {{Derivative
  expansion in quantum theory of gravitation}},}\ }\href {\doibase
  10.1143/PTP.93.615} {\bibfield  {journal} {\bibinfo  {journal} {Prog. Theor.
  Phys.}\ }\textbf {\bibinfo {volume} {93}},\ \bibinfo {pages} {615--620}
  (\bibinfo {year} {1995})},\ \Eprint {http://arxiv.org/abs/hep-th/9410111}
  {arXiv:hep-th/9410111 [hep-th]} \BibitemShut {NoStop}%
\bibitem [{\citenamefont {Alvarez}\ \emph {et~al.}(2006)\citenamefont
  {Alvarez}, \citenamefont {Blas}, \citenamefont {Garriga},\ and\ \citenamefont
  {Verdaguer}}]{Alvarez2006}%
  \BibitemOpen
  \bibfield  {author} {\bibinfo {author} {\bibfnamefont {E.}~\bibnamefont
  {Alvarez}}, \bibinfo {author} {\bibfnamefont {D.}~\bibnamefont {Blas}},
  \bibinfo {author} {\bibfnamefont {J.}~\bibnamefont {Garriga}}, \ and\
  \bibinfo {author} {\bibfnamefont {E.}~\bibnamefont {Verdaguer}},\ }\bibfield
  {title} {\enquote {\bibinfo {title} {{Transverse Fierz-Pauli symmetry}},}\
  }\href {\doibase 10.1016/j.nuclphysb.2006.08.003} {\bibfield  {journal}
  {\bibinfo  {journal} {Nucl. Phys.}\ }\textbf {\bibinfo {volume} {B756}},\
  \bibinfo {pages} {148--170} (\bibinfo {year} {2006})},\ \Eprint
  {http://arxiv.org/abs/hep-th/0606019} {arXiv:hep-th/0606019 [hep-th]}
  \BibitemShut {NoStop}%
\bibitem [{\citenamefont {Smolin}(2011)}]{Smolin2010}%
  \BibitemOpen
  \bibfield  {author} {\bibinfo {author} {\bibfnamefont {L.}~\bibnamefont
  {Smolin}},\ }\bibfield  {title} {\enquote {\bibinfo {title} {{Unimodular loop
  quantum gravity and the problems of time}},}\ }\href {\doibase
  10.1103/PhysRevD.84.044047} {\bibfield  {journal} {\bibinfo  {journal} {Phys.
  Rev.}\ }\textbf {\bibinfo {volume} {D84}},\ \bibinfo {pages} {044047}
  (\bibinfo {year} {2011})},\ \Eprint {http://arxiv.org/abs/1008.1759}
  {arXiv:1008.1759 [hep-th]} \BibitemShut {NoStop}%
\bibitem [{\citenamefont {Barcel\'o}\ \emph {et~al.}(2014)\citenamefont
  {Barcel\'o}, \citenamefont {Carballo-Rubio},\ and\ \citenamefont
  {Garay}}]{Barcelo2014}%
  \BibitemOpen
  \bibfield  {author} {\bibinfo {author} {\bibfnamefont {C.}~\bibnamefont
  {Barcel\'o}}, \bibinfo {author} {\bibfnamefont {R.}~\bibnamefont
  {Carballo-Rubio}}, \ and\ \bibinfo {author} {\bibfnamefont {L.~J.}\
  \bibnamefont {Garay}},\ }\bibfield  {title} {\enquote {\bibinfo {title}
  {{Unimodular gravity and general relativity from graviton
  self-interactions}},}\ }\href {\doibase 10.1103/PhysRevD.89.124019}
  {\bibfield  {journal} {\bibinfo  {journal} {Phys.Rev.}\ }\textbf {\bibinfo
  {volume} {D89}},\ \bibinfo {pages} {124019} (\bibinfo {year} {2014})},\
  \Eprint {http://arxiv.org/abs/1401.2941} {arXiv:1401.2941 [gr-qc]}
  \BibitemShut {NoStop}%
\bibitem [{\citenamefont {Herrero-Valea}(2018)}]{Herrero-Valea2018}%
  \BibitemOpen
  \bibfield  {author} {\bibinfo {author} {\bibfnamefont {M.}~\bibnamefont
  {Herrero-Valea}},\ }\bibfield  {title} {\enquote {\bibinfo {title} {{What do
  gravitons say about (unimodular) gravity?}}}\ }\href@noop {} {\  (\bibinfo
  {year} {2018})},\ \Eprint {http://arxiv.org/abs/1806.01869} {arXiv:1806.01869
  [hep-th]} \BibitemShut {NoStop}%
\bibitem [{\citenamefont {{Smolin}}(2009)}]{Smolin2009}%
  \BibitemOpen
  \bibfield  {author} {\bibinfo {author} {\bibfnamefont {L.}~\bibnamefont
  {{Smolin}}},\ }\bibfield  {title} {\enquote {\bibinfo {title} {{Quantization
  of unimodular gravity and the cosmological constant problems}},}\ }\href
  {\doibase 10.1103/PhysRevD.80.084003} {\bibfield  {journal} {\bibinfo
  {journal} {Physical Review D}\ }\textbf {\bibinfo {volume} {80}},\ \bibinfo
  {eid} {084003} (\bibinfo {year} {2009})},\ \Eprint
  {http://arxiv.org/abs/0904.4841} {arXiv:0904.4841 [hep-th]} \BibitemShut
  {NoStop}%
\bibitem [{\citenamefont {{Ellis}}\ \emph {et~al.}(2011)\citenamefont
  {{Ellis}}, \citenamefont {{van Elst}}, \citenamefont {{Murugan}},\ and\
  \citenamefont {{Uzan}}}]{Ellis2011}%
  \BibitemOpen
  \bibfield  {author} {\bibinfo {author} {\bibfnamefont {G.~F.~R.}\
  \bibnamefont {{Ellis}}}, \bibinfo {author} {\bibfnamefont {H.}~\bibnamefont
  {{van Elst}}}, \bibinfo {author} {\bibfnamefont {J.}~\bibnamefont
  {{Murugan}}}, \ and\ \bibinfo {author} {\bibfnamefont {J.-P.}\ \bibnamefont
  {{Uzan}}},\ }\bibfield  {title} {\enquote {\bibinfo {title} {{On the
  trace-free Einstein equations as a viable alternative to general
  relativity}},}\ }\href {\doibase 10.1088/0264-9381/28/22/225007} {\bibfield
  {journal} {\bibinfo  {journal} {Classical and Quantum Gravity}\ }\textbf
  {\bibinfo {volume} {28}},\ \bibinfo {eid} {225007} (\bibinfo {year}
  {2011})},\ \Eprint {http://arxiv.org/abs/1008.1196} {arXiv:1008.1196 [gr-qc]}
  \BibitemShut {NoStop}%
\bibitem [{\citenamefont {Ellis}(2014)}]{Ellis2013}%
  \BibitemOpen
  \bibfield  {author} {\bibinfo {author} {\bibfnamefont {G.~F.~R.}\
  \bibnamefont {Ellis}},\ }\bibfield  {title} {\enquote {\bibinfo {title} {{The
  Trace-Free Einstein Equations and inflation}},}\ }\href {\doibase
  10.1007/s10714-013-1619-5} {\bibfield  {journal} {\bibinfo  {journal}
  {Gen.Rel.Grav.}\ }\textbf {\bibinfo {volume} {46}},\ \bibinfo {pages} {1619}
  (\bibinfo {year} {2014})},\ \Eprint {http://arxiv.org/abs/1306.3021}
  {arXiv:1306.3021 [gr-qc]} \BibitemShut {NoStop}%
\bibitem [{\citenamefont {Barceló}\ \emph {et~al.}(2018)\citenamefont
  {Barceló}, \citenamefont {Carballo-Rubio},\ and\ \citenamefont
  {Garay}}]{BCRG2014}%
  \BibitemOpen
  \bibfield  {author} {\bibinfo {author} {\bibfnamefont {C.}~\bibnamefont
  {Barceló}}, \bibinfo {author} {\bibfnamefont {R.}~\bibnamefont
  {Carballo-Rubio}}, \ and\ \bibinfo {author} {\bibfnamefont {L.~J.}\
  \bibnamefont {Garay}},\ }\bibfield  {title} {\enquote {\bibinfo {title}
  {{Absence of cosmological constant problem in special relativistic field
  theory of gravity}},}\ }\href {\doibase 10.1016/j.aop.2018.08.016} {\bibfield
   {journal} {\bibinfo  {journal} {Annals Phys.}\ }\textbf {\bibinfo {volume}
  {398}},\ \bibinfo {pages} {9--23} (\bibinfo {year} {2018})},\ \Eprint
  {http://arxiv.org/abs/1406.7713} {arXiv:1406.7713 [gr-qc]} \BibitemShut
  {NoStop}%
\bibitem [{\citenamefont {Carballo-Rubio}(2015)}]{Carballo-Rubio2015}%
  \BibitemOpen
  \bibfield  {author} {\bibinfo {author} {\bibfnamefont {R.}~\bibnamefont
  {Carballo-Rubio}},\ }\bibfield  {title} {\enquote {\bibinfo {title}
  {{Longitudinal diffeomorphisms obstruct the protection of vacuum energy}},}\
  }\href {\doibase 10.1103/PhysRevD.91.124071} {\bibfield  {journal} {\bibinfo
  {journal} {Phys. Rev.}\ }\textbf {\bibinfo {volume} {D91}},\ \bibinfo {pages}
  {124071} (\bibinfo {year} {2015})},\ \Eprint
  {http://arxiv.org/abs/1502.05278} {arXiv:1502.05278 [gr-qc]} \BibitemShut
  {NoStop}%
\bibitem [{\citenamefont {Barceló}\ \emph {et~al.}(2015)\citenamefont
  {Barceló}, \citenamefont {Carballo-Rubio},\ and\ \citenamefont
  {Garay}}]{Barcelo2015}%
  \BibitemOpen
  \bibfield  {author} {\bibinfo {author} {\bibfnamefont {C.}~\bibnamefont
  {Barceló}}, \bibinfo {author} {\bibfnamefont {R.}~\bibnamefont
  {Carballo-Rubio}}, \ and\ \bibinfo {author} {\bibfnamefont {L.~J.}\
  \bibnamefont {Garay}},\ }\bibfield  {title} {\enquote {\bibinfo {title}
  {{Uncovering the effective spacetime: Lessons from the effective field theory
  rationale}},}\ }\href {\doibase 10.1142/S0218271815440198} {\bibfield
  {journal} {\bibinfo  {journal} {Int. J. Mod. Phys.}\ }\textbf {\bibinfo
  {volume} {D24}},\ \bibinfo {pages} {1544019} (\bibinfo {year} {2015})},\
  \Eprint {http://arxiv.org/abs/1505.05315} {arXiv:1505.05315 [hep-th]}
  \BibitemShut {NoStop}%
\bibitem [{\citenamefont {de~León~Ardón}\ \emph {et~al.}(2018)\citenamefont
  {de~León~Ardón}, \citenamefont {Ohta},\ and\ \citenamefont
  {Percacci}}]{Ardon2017}%
  \BibitemOpen
  \bibfield  {author} {\bibinfo {author} {\bibfnamefont {R.}~\bibnamefont
  {de~León~Ardón}}, \bibinfo {author} {\bibfnamefont {N.}~\bibnamefont {Ohta}},
  \ and\ \bibinfo {author} {\bibfnamefont {R.}~\bibnamefont {Percacci}},\
  }\bibfield  {title} {\enquote {\bibinfo {title} {{Path integral of unimodular
  gravity}},}\ }\href {\doibase 10.1103/PhysRevD.97.026007} {\bibfield
  {journal} {\bibinfo  {journal} {Phys. Rev.}\ }\textbf {\bibinfo {volume}
  {D97}},\ \bibinfo {pages} {026007} (\bibinfo {year} {2018})},\ \Eprint
  {http://arxiv.org/abs/1710.02457} {arXiv:1710.02457 [gr-qc]} \BibitemShut
  {NoStop}%
\bibitem [{\citenamefont {Nojiri}(2016)}]{Nojiri2016}%
  \BibitemOpen
  \bibfield  {author} {\bibinfo {author} {\bibfnamefont {S.}~\bibnamefont
  {Nojiri}},\ }\bibfield  {title} {\enquote {\bibinfo {title} {{Some solutions
  for one of the cosmological constant problems}},}\ }\href {\doibase
  10.1142/S0217732316502138} {\bibfield  {journal} {\bibinfo  {journal} {Mod.
  Phys. Lett.}\ }\textbf {\bibinfo {volume} {A31}},\ \bibinfo {pages} {1650213}
  (\bibinfo {year} {2016})},\ \Eprint {http://arxiv.org/abs/1601.02203}
  {arXiv:1601.02203 [hep-th]} \BibitemShut {NoStop}%
\bibitem [{\citenamefont {Mori}\ \emph {et~al.}(2017)\citenamefont {Mori},
  \citenamefont {Nitta},\ and\ \citenamefont {Nojiri}}]{Mori2017}%
  \BibitemOpen
  \bibfield  {author} {\bibinfo {author} {\bibfnamefont {T.}~\bibnamefont
  {Mori}}, \bibinfo {author} {\bibfnamefont {D.}~\bibnamefont {Nitta}}, \ and\
  \bibinfo {author} {\bibfnamefont {S.}~\bibnamefont {Nojiri}},\ }\bibfield
  {title} {\enquote {\bibinfo {title} {{BRS structure of Simple Model of
  Cosmological Constant and Cosmology}},}\ }\href {\doibase
  10.1103/PhysRevD.96.024009} {\bibfield  {journal} {\bibinfo  {journal} {Phys.
  Rev.}\ }\textbf {\bibinfo {volume} {D96}},\ \bibinfo {pages} {024009}
  (\bibinfo {year} {2017})},\ \Eprint {http://arxiv.org/abs/1702.07063}
  {arXiv:1702.07063 [hep-th]} \BibitemShut {NoStop}%
\bibitem [{\citenamefont {Nojiri}(2018)}]{Nojiri2018}%
  \BibitemOpen
  \bibfield  {author} {\bibinfo {author} {\bibfnamefont {S.}~\bibnamefont
  {Nojiri}},\ }\bibfield  {title} {\enquote {\bibinfo {title} {{Cosmological
  constant and renormalization of gravity}},}\ }\bibfield  {booktitle} {\emph
  {\bibinfo {booktitle} {{Proceedings, 4th Workshop on Cosmology and the
  Quantum Vacuum: Segovia, Spain, September 4-8, 2017}}},\ }\href {\doibase
  10.3390/galaxies6010024} {\bibfield  {journal} {\bibinfo  {journal}
  {Galaxies}\ }\textbf {\bibinfo {volume} {6}},\ \bibinfo {pages} {24}
  (\bibinfo {year} {2018})},\ \Eprint {http://arxiv.org/abs/1802.04596}
  {arXiv:1802.04596 [hep-th]} \BibitemShut {NoStop}%
\bibitem [{\citenamefont {Mori}\ and\ \citenamefont {Nojiri}(2018)}]{Mori2018}%
  \BibitemOpen
  \bibfield  {author} {\bibinfo {author} {\bibfnamefont {T.}~\bibnamefont
  {Mori}}\ and\ \bibinfo {author} {\bibfnamefont {S.}~\bibnamefont {Nojiri}},\
  }\bibfield  {title} {\enquote {\bibinfo {title} {{Topological Gravity
  motivated by Renormalization Group}},}\ }\href@noop {} {\  (\bibinfo {year}
  {2018})},\ \Eprint {http://arxiv.org/abs/1809.02344} {arXiv:1809.02344
  [hep-th]} \BibitemShut {NoStop}%
\bibitem [{\citenamefont {Horowitz}(1989)}]{Horowitz1989}%
  \BibitemOpen
  \bibfield  {author} {\bibinfo {author} {\bibfnamefont {G.~T.}\ \bibnamefont
  {Horowitz}},\ }\bibfield  {title} {\enquote {\bibinfo {title} {{Exactly
  Soluble Diffeomorphism Invariant Theories}},}\ }\href {\doibase
  10.1007/BF01218410} {\bibfield  {journal} {\bibinfo  {journal} {Commun. Math.
  Phys.}\ }\textbf {\bibinfo {volume} {125}},\ \bibinfo {pages} {417} (\bibinfo
  {year} {1989})}\BibitemShut {NoStop}%
\bibitem [{\citenamefont {Baez}(1996)}]{Baez1995}%
  \BibitemOpen
  \bibfield  {author} {\bibinfo {author} {\bibfnamefont {J.~C.}\ \bibnamefont
  {Baez}},\ }\bibfield  {title} {\enquote {\bibinfo {title} {{Four-Dimensional
  BF theory with cosmological term as a topological quantum field theory}},}\
  }\href {\doibase 10.1007/BF00398315} {\bibfield  {journal} {\bibinfo
  {journal} {Lett. Math. Phys.}\ }\textbf {\bibinfo {volume} {38}},\ \bibinfo
  {pages} {129--143} (\bibinfo {year} {1996})},\ \Eprint
  {http://arxiv.org/abs/q-alg/9507006} {arXiv:q-alg/9507006 [q-alg]}
  \BibitemShut {NoStop}%
\bibitem [{\citenamefont {Cattaneo}\ \emph {et~al.}(1995)\citenamefont
  {Cattaneo}, \citenamefont {Cotta-Ramusino}, \citenamefont {Frohlich},\ and\
  \citenamefont {Martellini}}]{Cattaneo1995}%
  \BibitemOpen
  \bibfield  {author} {\bibinfo {author} {\bibfnamefont {A.~S.}\ \bibnamefont
  {Cattaneo}}, \bibinfo {author} {\bibfnamefont {P.}~\bibnamefont
  {Cotta-Ramusino}}, \bibinfo {author} {\bibfnamefont {J.}~\bibnamefont
  {Frohlich}}, \ and\ \bibinfo {author} {\bibfnamefont {M.}~\bibnamefont
  {Martellini}},\ }\bibfield  {title} {\enquote {\bibinfo {title} {{Topological
  BF theories in three-dimensions and four-dimensions}},}\ }\href {\doibase
  10.1063/1.531238} {\bibfield  {journal} {\bibinfo  {journal} {J. Math.
  Phys.}\ }\textbf {\bibinfo {volume} {36}},\ \bibinfo {pages} {6137--6160}
  (\bibinfo {year} {1995})},\ \Eprint {http://arxiv.org/abs/hep-th/9505027}
  {arXiv:hep-th/9505027 [hep-th]} \BibitemShut {NoStop}%
\bibitem [{\citenamefont {Baez}(2000)}]{Baez1999}%
  \BibitemOpen
  \bibfield  {author} {\bibinfo {author} {\bibfnamefont {J.~C.}\ \bibnamefont
  {Baez}},\ }\bibfield  {title} {\enquote {\bibinfo {title} {{An Introduction
  to spin foam models of quantum gravity and BF theory}},}\ }\bibfield
  {booktitle} {\emph {\bibinfo {booktitle} {{Geometry and quantum physics.
  Proceedings, 38. Internationale Universitätswochen für Kern- und
  Teilchenphysik, IUKT 38: Schladming, Austria, January 9-16, 1999}}},\ }\href
  {\doibase 10.1007/3-540-46552-9_2} {\bibfield  {journal} {\bibinfo  {journal}
  {Lect. Notes Phys.}\ }\textbf {\bibinfo {volume} {543}},\ \bibinfo {pages}
  {25--94} (\bibinfo {year} {2000})},\ \Eprint
  {http://arxiv.org/abs/gr-qc/9905087} {arXiv:gr-qc/9905087 [gr-qc]}
  \BibitemShut {NoStop}%
\bibitem [{\citenamefont {Gielen}\ and\ \citenamefont
  {Oriti}(2010)}]{Gielen2010}%
  \BibitemOpen
  \bibfield  {author} {\bibinfo {author} {\bibfnamefont {S.}~\bibnamefont
  {Gielen}}\ and\ \bibinfo {author} {\bibfnamefont {D.}~\bibnamefont {Oriti}},\
  }\bibfield  {title} {\enquote {\bibinfo {title} {{Classical general
  relativity as BF-Plebanski theory with linear constraints}},}\ }\href
  {\doibase 10.1088/0264-9381/27/18/185017} {\bibfield  {journal} {\bibinfo
  {journal} {Class. Quant. Grav.}\ }\textbf {\bibinfo {volume} {27}},\ \bibinfo
  {pages} {185017} (\bibinfo {year} {2010})},\ \Eprint
  {http://arxiv.org/abs/1004.5371} {arXiv:1004.5371 [gr-qc]} \BibitemShut
  {NoStop}%
\bibitem [{\citenamefont {Freidel}\ and\ \citenamefont
  {Speziale}(2012)}]{Freidel2012}%
  \BibitemOpen
  \bibfield  {author} {\bibinfo {author} {\bibfnamefont {L.}~\bibnamefont
  {Freidel}}\ and\ \bibinfo {author} {\bibfnamefont {S.}~\bibnamefont
  {Speziale}},\ }\bibfield  {title} {\enquote {\bibinfo {title} {{On the
  relations between gravity and BF theories}},}\ }\href {\doibase
  10.3842/SIGMA.2012.032} {\bibfield  {journal} {\bibinfo  {journal} {SIGMA}\
  }\textbf {\bibinfo {volume} {8}},\ \bibinfo {pages} {032} (\bibinfo {year}
  {2012})},\ \Eprint {http://arxiv.org/abs/1201.4247} {arXiv:1201.4247 [gr-qc]}
  \BibitemShut {NoStop}%
\bibitem [{\citenamefont {Celada}\ \emph {et~al.}(2016)\citenamefont {Celada},
  \citenamefont {González},\ and\ \citenamefont {Montesinos}}]{Celada2016}%
  \BibitemOpen
  \bibfield  {author} {\bibinfo {author} {\bibfnamefont {M.}~\bibnamefont
  {Celada}}, \bibinfo {author} {\bibfnamefont {D.}~\bibnamefont {González}}, \
  and\ \bibinfo {author} {\bibfnamefont {M.}~\bibnamefont {Montesinos}},\
  }\bibfield  {title} {\enquote {\bibinfo {title} {{$BF$ gravity}},}\ }\href
  {\doibase 10.1088/0264-9381/33/21/213001} {\bibfield  {journal} {\bibinfo
  {journal} {Class. Quant. Grav.}\ }\textbf {\bibinfo {volume} {33}},\ \bibinfo
  {pages} {213001} (\bibinfo {year} {2016})},\ \Eprint
  {http://arxiv.org/abs/1610.02020} {arXiv:1610.02020 [gr-qc]} \BibitemShut
  {NoStop}%
\bibitem [{\citenamefont {de~Gracia}\ \emph {et~al.}(2017)\citenamefont
  {de~Gracia}, \citenamefont {Pimentel},\ and\ \citenamefont
  {Valcárcel}}]{deGracia2017}%
  \BibitemOpen
  \bibfield  {author} {\bibinfo {author} {\bibfnamefont {G.~B.}\ \bibnamefont
  {de~Gracia}}, \bibinfo {author} {\bibfnamefont {B.~M.}\ \bibnamefont
  {Pimentel}}, \ and\ \bibinfo {author} {\bibfnamefont {C.~E.}\ \bibnamefont
  {Valcárcel}},\ }\bibfield  {title} {\enquote {\bibinfo {title}
  {{Hamilton-Jacobi analysis of the four dimensional BF model with cosmological
  term}},}\ }\href@noop {} {\  (\bibinfo {year} {2017})},\ \Eprint
  {http://arxiv.org/abs/1702.00863} {arXiv:1702.00863 [hep-th]} \BibitemShut
  {NoStop}%
\bibitem [{\citenamefont {Bradonjic}\ and\ \citenamefont
  {Stachel}(2012)}]{Bradonjic2011}%
  \BibitemOpen
  \bibfield  {author} {\bibinfo {author} {\bibfnamefont {K.}~\bibnamefont
  {Bradonjic}}\ and\ \bibinfo {author} {\bibfnamefont {J.}~\bibnamefont
  {Stachel}},\ }\bibfield  {title} {\enquote {\bibinfo {title} {{Unimodular
  Conformal and Projective Relativity}},}\ }\href {\doibase
  10.1209/0295-5075/97/10001} {\bibfield  {journal} {\bibinfo  {journal}
  {Europhys. Lett.}\ }\textbf {\bibinfo {volume} {97}},\ \bibinfo {pages}
  {10001} (\bibinfo {year} {2012})},\ \Eprint {http://arxiv.org/abs/1110.2159}
  {arXiv:1110.2159 [gr-qc]} \BibitemShut {NoStop}%
\bibitem [{\citenamefont {Bradonjic}(2014)}]{Bradonjic2014}%
  \BibitemOpen
  \bibfield  {author} {\bibinfo {author} {\bibfnamefont {K.}~\bibnamefont
  {Bradonjic}},\ }\bibfield  {title} {\enquote {\bibinfo {title} {{Unimodular
  Conformal and Projective Relativity: An Illustrated Introduction}},}\
  }\bibfield  {booktitle} {\emph {\bibinfo {booktitle} {{Proceedings, 12th
  International Symposium on Frontiers of Fundamental Physics and Physics
  Education Research (FFP12): Udine, Italy, November 21-23, 2011}}},\ }\href
  {\doibase 10.1007/978-3-319-00297-2_20} {\bibfield  {journal} {\bibinfo
  {journal} {Springer Proc. Phys.}\ }\textbf {\bibinfo {volume} {145}},\
  \bibinfo {pages} {197--203} (\bibinfo {year} {2014})}\BibitemShut {NoStop}%
\bibitem [{\citenamefont {Salim}\ and\ \citenamefont
  {Sautu}(1996)}]{Salim1996}%
  \BibitemOpen
  \bibfield  {author} {\bibinfo {author} {\bibfnamefont {J.~M.}\ \bibnamefont
  {Salim}}\ and\ \bibinfo {author} {\bibfnamefont {S.~L.}\ \bibnamefont
  {Sautu}},\ }\bibfield  {title} {\enquote {\bibinfo {title} {{Gravitational
  theory in Weyl integrable space-time}},}\ }\href {\doibase
  10.1088/0264-9381/13/3/004} {\bibfield  {journal} {\bibinfo  {journal}
  {Class. Quant. Grav.}\ }\textbf {\bibinfo {volume} {13}},\ \bibinfo {pages}
  {353--360} (\bibinfo {year} {1996})}\BibitemShut {NoStop}%
\bibitem [{\citenamefont {Romero}\ \emph {et~al.}(2012)\citenamefont {Romero},
  \citenamefont {Fonseca-Neto},\ and\ \citenamefont {Pucheu}}]{Romero2012}%
  \BibitemOpen
  \bibfield  {author} {\bibinfo {author} {\bibfnamefont {C.}~\bibnamefont
  {Romero}}, \bibinfo {author} {\bibfnamefont {J.~B.}\ \bibnamefont
  {Fonseca-Neto}}, \ and\ \bibinfo {author} {\bibfnamefont {M.~L.}\
  \bibnamefont {Pucheu}},\ }\bibfield  {title} {\enquote {\bibinfo {title}
  {{General Relativity and Weyl Geometry}},}\ }\href {\doibase
  10.1088/0264-9381/29/15/155015} {\bibfield  {journal} {\bibinfo  {journal}
  {Class. Quant. Grav.}\ }\textbf {\bibinfo {volume} {29}},\ \bibinfo {pages}
  {155015} (\bibinfo {year} {2012})},\ \Eprint {http://arxiv.org/abs/1201.1469}
  {arXiv:1201.1469 [gr-qc]} \BibitemShut {NoStop}%
\bibitem [{\citenamefont {Yuan}\ and\ \citenamefont
  {Huang}(2013)}]{YuanHuang2013}%
  \BibitemOpen
  \bibfield  {author} {\bibinfo {author} {\bibfnamefont {F.~F.}\ \bibnamefont
  {Yuan}}\ and\ \bibinfo {author} {\bibfnamefont {Y.~C.}\ \bibnamefont
  {Huang}},\ }\bibfield  {title} {\enquote {\bibinfo {title} {{A modified
  variational principle for gravity in the modified Weyl geometry}},}\ }\href
  {\doibase 10.1088/0264-9381/30/19/195008} {\bibfield  {journal} {\bibinfo
  {journal} {Class. Quant. Grav.}\ }\textbf {\bibinfo {volume} {30}},\ \bibinfo
  {pages} {195008} (\bibinfo {year} {2013})},\ \Eprint
  {http://arxiv.org/abs/1301.1316} {arXiv:1301.1316 [gr-qc]} \BibitemShut
  {NoStop}%
\bibitem [{\citenamefont {Barceló}\ \emph {et~al.}(2017)\citenamefont
  {Barceló}, \citenamefont {Carballo-Rubio},\ and\ \citenamefont
  {Garay}}]{Barcelo2017}%
  \BibitemOpen
  \bibfield  {author} {\bibinfo {author} {\bibfnamefont {C.}~\bibnamefont
  {Barceló}}, \bibinfo {author} {\bibfnamefont {R.}~\bibnamefont
  {Carballo-Rubio}}, \ and\ \bibinfo {author} {\bibfnamefont {L.~J.}\
  \bibnamefont {Garay}},\ }\bibfield  {title} {\enquote {\bibinfo {title}
  {{Weyl relativity: A novel approach to Weyl's ideas}},}\ }\href {\doibase
  10.1088/1475-7516/2017/06/014} {\bibfield  {journal} {\bibinfo  {journal}
  {JCAP}\ }\textbf {\bibinfo {volume} {1706}},\ \bibinfo {pages} {014}
  (\bibinfo {year} {2017})},\ \Eprint {http://arxiv.org/abs/1703.06355}
  {arXiv:1703.06355 [gr-qc]} \BibitemShut {NoStop}%
\bibitem [{\citenamefont {Birrell}\ and\ \citenamefont
  {Davies}(1984)}]{Birrell1982}%
  \BibitemOpen
  \bibfield  {author} {\bibinfo {author} {\bibfnamefont {N.~D.}\ \bibnamefont
  {Birrell}}\ and\ \bibinfo {author} {\bibfnamefont {P.~C.~W.}\ \bibnamefont
  {Davies}},\ }\href {\doibase 10.1017/CBO9780511622632} {\emph {\bibinfo
  {title} {{Quantum Fields in Curved Space}}}},\ Cambridge Monographs on
  Mathematical Physics\ (\bibinfo  {publisher} {Cambridge Univ. Press},\
  \bibinfo {address} {Cambridge, UK},\ \bibinfo {year} {1984})\BibitemShut
  {NoStop}%
\bibitem [{\citenamefont {{Visser}}(2002)}]{Visser2002}%
  \BibitemOpen
  \bibfield  {author} {\bibinfo {author} {\bibfnamefont {M.}~\bibnamefont
  {{Visser}}},\ }\bibfield  {title} {\enquote {\bibinfo {title} {{Sakharov's
  Induced Gravity}},}\ }\href {\doibase 10.1142/S0217732302006886} {\bibfield
  {journal} {\bibinfo  {journal} {Modern Physics Letters A}\ }\textbf {\bibinfo
  {volume} {17}},\ \bibinfo {pages} {977--991} (\bibinfo {year} {2002})},\
  \Eprint {http://arxiv.org/abs/gr-qc/0204062} {gr-qc/0204062} \BibitemShut
  {NoStop}%
\bibitem [{\citenamefont {{Vassilevich}}(2003)}]{Vassilevich2003}%
  \BibitemOpen
  \bibfield  {author} {\bibinfo {author} {\bibfnamefont {D.~V.}\ \bibnamefont
  {{Vassilevich}}},\ }\bibfield  {title} {\enquote {\bibinfo {title} {{Heat
  kernel expansion: user's manual}},}\ }\href {\doibase
  10.1016/j.physrep.2003.09.002} {\bibfield  {journal} {\bibinfo  {journal}
  {Phys. Rep.}\ }\textbf {\bibinfo {volume} {388}},\ \bibinfo {pages}
  {279--360} (\bibinfo {year} {2003})},\ \Eprint
  {http://arxiv.org/abs/hep-th/0306138} {hep-th/0306138} \BibitemShut {NoStop}%
\bibitem [{\citenamefont {Mukhanov}\ and\ \citenamefont
  {Winitzki}(2007)}]{Mukhanov2007}%
  \BibitemOpen
  \bibfield  {author} {\bibinfo {author} {\bibfnamefont {V.}~\bibnamefont
  {Mukhanov}}\ and\ \bibinfo {author} {\bibfnamefont {S.}~\bibnamefont
  {Winitzki}},\ }\href {http://books.google.es/books?id=vmwHoxf2958C} {\emph
  {\bibinfo {title} {Introduction to Quantum Effects in Gravity}}}\ (\bibinfo
  {publisher} {Cambridge University Press},\ \bibinfo {year}
  {2007})\BibitemShut {NoStop}%
\bibitem [{\citenamefont {Martin}(2012)}]{Martin2012}%
  \BibitemOpen
  \bibfield  {author} {\bibinfo {author} {\bibfnamefont {J.}~\bibnamefont
  {Martin}},\ }\bibfield  {title} {\enquote {\bibinfo {title} {{Everything You
  Always Wanted To Know About The Cosmological Constant Problem (But Were
  Afraid To Ask)}},}\ }\href {\doibase 10.1016/j.crhy.2012.04.008} {\bibfield
  {journal} {\bibinfo  {journal} {Comptes Rendus Physique}\ }\textbf {\bibinfo
  {volume} {13}},\ \bibinfo {pages} {566--665} (\bibinfo {year} {2012})},\
  \Eprint {http://arxiv.org/abs/1205.3365} {arXiv:1205.3365 [astro-ph.CO]}
  \BibitemShut {NoStop}%
\bibitem [{\citenamefont {Josset}\ \emph {et~al.}(2017)\citenamefont {Josset},
  \citenamefont {Perez},\ and\ \citenamefont {Sudarsky}}]{Josset2016}%
  \BibitemOpen
  \bibfield  {author} {\bibinfo {author} {\bibfnamefont {T.}~\bibnamefont
  {Josset}}, \bibinfo {author} {\bibfnamefont {A.}~\bibnamefont {Perez}}, \
  and\ \bibinfo {author} {\bibfnamefont {D.}~\bibnamefont {Sudarsky}},\
  }\bibfield  {title} {\enquote {\bibinfo {title} {{Dark Energy from Violation
  of Energy Conservation}},}\ }\href {\doibase 10.1103/PhysRevLett.118.021102}
  {\bibfield  {journal} {\bibinfo  {journal} {Phys. Rev. Lett.}\ }\textbf
  {\bibinfo {volume} {118}},\ \bibinfo {pages} {021102} (\bibinfo {year}
  {2017})},\ \Eprint {http://arxiv.org/abs/1604.04183} {arXiv:1604.04183
  [gr-qc]} \BibitemShut {NoStop}%
\bibitem [{\citenamefont {Perez}\ and\ \citenamefont
  {Sudarsky}(2017)}]{Perez2017}%
  \BibitemOpen
  \bibfield  {author} {\bibinfo {author} {\bibfnamefont {A.}~\bibnamefont
  {Perez}}\ and\ \bibinfo {author} {\bibfnamefont {D.}~\bibnamefont
  {Sudarsky}},\ }\bibfield  {title} {\enquote {\bibinfo {title} {{Dark energy
  from quantum gravity discreteness}},}\ }\href@noop {} {\  (\bibinfo {year}
  {2017})},\ \Eprint {http://arxiv.org/abs/1711.05183} {arXiv:1711.05183
  [gr-qc]} \BibitemShut {NoStop}%
\bibitem [{\citenamefont {Perez}\ \emph {et~al.}(2018)\citenamefont {Perez},
  \citenamefont {Sudarsky},\ and\ \citenamefont {Bjorken}}]{Perez2018}%
  \BibitemOpen
  \bibfield  {author} {\bibinfo {author} {\bibfnamefont {A.}~\bibnamefont
  {Perez}}, \bibinfo {author} {\bibfnamefont {D.}~\bibnamefont {Sudarsky}}, \
  and\ \bibinfo {author} {\bibfnamefont {J.~D.}\ \bibnamefont {Bjorken}},\
  }\bibfield  {title} {\enquote {\bibinfo {title} {{A microscopic model for an
  emergent cosmological constant}},}\ }\href@noop {} {\  (\bibinfo {year}
  {2018})},\ \Eprint {http://arxiv.org/abs/1804.07162} {arXiv:1804.07162
  [gr-qc]} \BibitemShut {NoStop}%
\bibitem [{\citenamefont {Volovik}(2006{\natexlab{a}})}]{Volovik2003}%
  \BibitemOpen
  \bibfield  {author} {\bibinfo {author} {\bibfnamefont {G.~E.}\ \bibnamefont
  {Volovik}},\ }\bibfield  {title} {\enquote {\bibinfo {title} {{The Universe
  in a helium droplet}},}\ }\href@noop {} {\bibfield  {journal} {\bibinfo
  {journal} {Int. Ser. Monogr. Phys.}\ }\textbf {\bibinfo {volume} {117}},\
  \bibinfo {pages} {1--526} (\bibinfo {year} {2006}{\natexlab{a}})}\BibitemShut
  {NoStop}%
\bibitem [{\citenamefont {Volovik}(2005)}]{Volovik2004}%
  \BibitemOpen
  \bibfield  {author} {\bibinfo {author} {\bibfnamefont {G.~E.}\ \bibnamefont
  {Volovik}},\ }\bibfield  {title} {\enquote {\bibinfo {title} {{Cosmological
  constant and vacuum energy}},}\ }\href {\doibase 10.1002/andp.200410123}
  {\bibfield  {journal} {\bibinfo  {journal} {Annalen Phys.}\ }\textbf
  {\bibinfo {volume} {14}},\ \bibinfo {pages} {165--176} (\bibinfo {year}
  {2005})},\ \Eprint {http://arxiv.org/abs/gr-qc/0405012} {arXiv:gr-qc/0405012
  [gr-qc]} \BibitemShut {NoStop}%
\bibitem [{\citenamefont {Volovik}(2006{\natexlab{b}})}]{Volovik2006}%
  \BibitemOpen
  \bibfield  {author} {\bibinfo {author} {\bibfnamefont {G.~E.}\ \bibnamefont
  {Volovik}},\ }\bibfield  {title} {\enquote {\bibinfo {title} {{Vacuum Energy:
  Myths and Reality}},}\ }\href {\doibase 10.1142/S0218271806009431} {\bibfield
   {journal} {\bibinfo  {journal} {Int. J. Mod. Phys.}\ }\textbf {\bibinfo
  {volume} {D15}},\ \bibinfo {pages} {1987--2010} (\bibinfo {year}
  {2006}{\natexlab{b}})},\ \Eprint {http://arxiv.org/abs/gr-qc/0604062}
  {arXiv:gr-qc/0604062 [gr-qc]} \BibitemShut {NoStop}%
\bibitem [{\citenamefont {Klinkhamer}(2008)}]{Klinkhamer2008}%
  \BibitemOpen
  \bibfield  {author} {\bibinfo {author} {\bibfnamefont {F.~R.}\ \bibnamefont
  {Klinkhamer}},\ }\bibfield  {title} {\enquote {\bibinfo {title} {{Equilibrium
  boundary conditions, dynamic vacuum energy, and the Big Bang}},}\ }\href
  {\doibase 10.1103/PhysRevD.78.083533} {\bibfield  {journal} {\bibinfo
  {journal} {Phys. Rev.}\ }\textbf {\bibinfo {volume} {D78}},\ \bibinfo {pages}
  {083533} (\bibinfo {year} {2008})},\ \Eprint {http://arxiv.org/abs/0803.0281}
  {arXiv:0803.0281 [gr-qc]} \BibitemShut {NoStop}%
\bibitem [{\citenamefont {Klinkhamer}\ and\ \citenamefont
  {Volovik}(2008)}]{Klinkhamer2008b}%
  \BibitemOpen
  \bibfield  {author} {\bibinfo {author} {\bibfnamefont {F.~R.}\ \bibnamefont
  {Klinkhamer}}\ and\ \bibinfo {author} {\bibfnamefont {G.~E.}\ \bibnamefont
  {Volovik}},\ }\bibfield  {title} {\enquote {\bibinfo {title} {{Dynamic vacuum
  variable and equilibrium approach in cosmology}},}\ }\href {\doibase
  10.1103/PhysRevD.78.063528} {\bibfield  {journal} {\bibinfo  {journal} {Phys.
  Rev.}\ }\textbf {\bibinfo {volume} {D78}},\ \bibinfo {pages} {063528}
  (\bibinfo {year} {2008})},\ \Eprint {http://arxiv.org/abs/0806.2805}
  {arXiv:0806.2805 [gr-qc]} \BibitemShut {NoStop}%
\bibitem [{\citenamefont {Klinkhamer}\ and\ \citenamefont
  {Volovik}(2009)}]{Klinkhamer2009}%
  \BibitemOpen
  \bibfield  {author} {\bibinfo {author} {\bibfnamefont {F.~R.}\ \bibnamefont
  {Klinkhamer}}\ and\ \bibinfo {author} {\bibfnamefont {G.~E.}\ \bibnamefont
  {Volovik}},\ }\bibfield  {title} {\enquote {\bibinfo {title} {{Gluonic
  vacuum, q-theory, and the cosmological constant}},}\ }\href {\doibase
  10.1103/PhysRevD.79.063527} {\bibfield  {journal} {\bibinfo  {journal} {Phys.
  Rev.}\ }\textbf {\bibinfo {volume} {D79}},\ \bibinfo {pages} {063527}
  (\bibinfo {year} {2009})},\ \Eprint {http://arxiv.org/abs/0811.4347}
  {arXiv:0811.4347 [gr-qc]} \BibitemShut {NoStop}%
\bibitem [{\citenamefont {Alexander}\ \emph {et~al.}(2018)\citenamefont
  {Alexander}, \citenamefont {Magueijo},\ and\ \citenamefont
  {Smolin}}]{Alexander2018}%
  \BibitemOpen
  \bibfield  {author} {\bibinfo {author} {\bibfnamefont {S.}~\bibnamefont
  {Alexander}}, \bibinfo {author} {\bibfnamefont {J.}~\bibnamefont {Magueijo}},
  \ and\ \bibinfo {author} {\bibfnamefont {L.}~\bibnamefont {Smolin}},\
  }\bibfield  {title} {\enquote {\bibinfo {title} {{The quantum cosmological
  constant}},}\ }\href@noop {} {\  (\bibinfo {year} {2018})},\ \Eprint
  {http://arxiv.org/abs/1807.01381} {arXiv:1807.01381 [gr-qc]} \BibitemShut
  {NoStop}%
\bibitem [{\citenamefont {Kaloper}\ and\ \citenamefont
  {Padilla}(2014{\natexlab{a}})}]{Kaloper2013}%
  \BibitemOpen
  \bibfield  {author} {\bibinfo {author} {\bibfnamefont {N.}~\bibnamefont
  {Kaloper}}\ and\ \bibinfo {author} {\bibfnamefont {A.}~\bibnamefont
  {Padilla}},\ }\bibfield  {title} {\enquote {\bibinfo {title} {{Sequestering
  the Standard Model Vacuum Energy}},}\ }\href {\doibase
  10.1103/PhysRevLett.112.091304} {\bibfield  {journal} {\bibinfo  {journal}
  {Phys. Rev. Lett.}\ }\textbf {\bibinfo {volume} {112}},\ \bibinfo {pages}
  {091304} (\bibinfo {year} {2014}{\natexlab{a}})},\ \Eprint
  {http://arxiv.org/abs/1309.6562} {arXiv:1309.6562 [hep-th]} \BibitemShut
  {NoStop}%
\bibitem [{\citenamefont {Kaloper}\ and\ \citenamefont
  {Padilla}(2014{\natexlab{b}})}]{Kaloper2014}%
  \BibitemOpen
  \bibfield  {author} {\bibinfo {author} {\bibfnamefont {N.}~\bibnamefont
  {Kaloper}}\ and\ \bibinfo {author} {\bibfnamefont {A.}~\bibnamefont
  {Padilla}},\ }\bibfield  {title} {\enquote {\bibinfo {title} {{Vacuum Energy
  Sequestering: The Framework and Its Cosmological Consequences}},}\ }\href
  {\doibase 10.1103/PhysRevD.90.084023, 10.1103/PhysRevD.90.109901} {\bibfield
  {journal} {\bibinfo  {journal} {Phys. Rev.}\ }\textbf {\bibinfo {volume}
  {D90}},\ \bibinfo {pages} {084023} (\bibinfo {year} {2014}{\natexlab{b}})},\
  \bibinfo {note} {[Addendum: Phys. Rev.D90,no.10,109901(2014)]},\ \Eprint
  {http://arxiv.org/abs/1406.0711} {arXiv:1406.0711 [hep-th]} \BibitemShut
  {NoStop}%
\bibitem [{\citenamefont {Tsukamoto}\ \emph {et~al.}(2017)\citenamefont
  {Tsukamoto}, \citenamefont {Katsuragawa},\ and\ \citenamefont
  {Nojiri}}]{Tsukamoto2017}%
  \BibitemOpen
  \bibfield  {author} {\bibinfo {author} {\bibfnamefont {T.}~\bibnamefont
  {Tsukamoto}}, \bibinfo {author} {\bibfnamefont {T.}~\bibnamefont
  {Katsuragawa}}, \ and\ \bibinfo {author} {\bibfnamefont {S.}~\bibnamefont
  {Nojiri}},\ }\bibfield  {title} {\enquote {\bibinfo {title} {{Sequestering
  mechanism in scalar-tensor gravity}},}\ }\href {\doibase
  10.1103/PhysRevD.96.124003} {\bibfield  {journal} {\bibinfo  {journal} {Phys.
  Rev.}\ }\textbf {\bibinfo {volume} {D96}},\ \bibinfo {pages} {124003}
  (\bibinfo {year} {2017})},\ \Eprint {http://arxiv.org/abs/1710.06086}
  {arXiv:1710.06086 [hep-th]} \BibitemShut {NoStop}%
\bibitem [{\citenamefont {Weinberg}(1989)}]{Weinberg1989}%
  \BibitemOpen
  \bibfield  {author} {\bibinfo {author} {\bibfnamefont {S.}~\bibnamefont
  {Weinberg}},\ }\bibfield  {title} {\enquote {\bibinfo {title} {The
  cosmological constant problem},}\ }\href {\doibase 10.1103/RevModPhys.61.1}
  {\bibfield  {journal} {\bibinfo  {journal} {Rev. Mod. Phys.}\ }\textbf
  {\bibinfo {volume} {61}},\ \bibinfo {pages} {1--23} (\bibinfo {year}
  {1989})}\BibitemShut {NoStop}%
\bibitem [{\citenamefont {Ledermann}(1993)}]{Ledermann1993}%
  \BibitemOpen
  \bibfield  {author} {\bibinfo {author} {\bibfnamefont {W.}~\bibnamefont
  {Ledermann}},\ }\bibfield  {title} {\enquote {\bibinfo {title} {A note on
  skew-symmetric determinants},}\ }\href {\doibase 10.1017/S0013091500018423}
  {\bibfield  {journal} {\bibinfo  {journal} {Proc. Edinburgh Math. Soc.}\
  }\textbf {\bibinfo {volume} {36}},\ \bibinfo {pages} {335--338} (\bibinfo
  {year} {1993})}\BibitemShut {NoStop}%
\end{thebibliography}%

\end{document}